\newcommand{\tw}{t_{\rm w}}

\newcommand{\Leff}{L_{\rm eff}}
\newcommand{\Lmax}{L_{\rm max}}
\newcommand{\Lo}{L_0}

\newcommand{\Lovlp}{L_{\rm \Delta T}}

\newcommand \be {\begin{equation}}
\newcommand \ee {\end{equation}}

\newcommand{\kb}{k_{\rm B}}

\documentclass[aps,prl,twocolumn,superscriptaddress]{revtex4}

\usepackage{graphicx}

\begin{document}

%\date{\today}

J{\"o}nsson, Yoshino and Nordblad reply: In their comment to our 
work \cite{letter}, L. Berthier and J. P. Bouchaud \cite{bb-comment} 
report results of temperature-shifts 
$\Delta T$ 
in a numerical simulation of a 4 dimensional EA Ising spin-glass model using the same 
thermal protocol as in \cite{letter}. Their scaling analysis 
yields apparently similar result to ours. 

In our experiments the data are limited to a narrow temperature 
region $|\Delta T| < 0.02 \, T_{\rm g}$. The expected saturation of 
$\Leff$ to the overlap length $\Lovlp$ \cite{droplet} is not observed for these small values of $|\Delta T|$. In this reply, we instead point out that our 
experimental results can be understood in terms 
of temperature chaos as  {\it partial} rejuvenation in a 
{\it weakly perturbed regime}.

Within the droplet theory \cite{droplet}, it is assumed that there are dangerously 
irrelevant droplets whose free-energy gaps $F_{L}$ are arbitrary small. 
The gap can be smaller than $dF$ with probability $\tilde{\rho}(0)dF/\Upsilon(L/\Lo)^{\theta}$, where
$\tilde{\rho}(0) >0$.
Consequently a perturbation which amounts to a small free-energy gain of 
order $\Delta U (L/\Lo)^{\alpha}$ can induce a droplet excitation with a probability
$p(L/L_{\Delta U}) \sim \tilde{\rho}(0)(L/L_{\Delta U})^{\zeta}$. 
Here $L_{\Delta U} = \Lo(\Delta U/\Upsilon)^{-1/\zeta}$ is the overlap
length with the chaos exponent $\zeta=\alpha-\theta$. In the case of 
T-shifts, $\Delta U= \kb |\Delta T|$ and $\alpha=d_{s}/2$,
where $d_{s}$ is the surface  fractal dimension of the droplets.
Importantly, the probability $p(L/L_{\Delta U})$ is non-zero 
in the weakly perturbed regime  $L < L_{\Delta U}$
and  {\it increases with $L$} if $\zeta>0$
smoothly connecting to the strongly perturbed regime 
$L > L_{\Delta U}$ where $p(L/L_{\Delta U})=1$ \cite{SY}.

\begin{figure}[t]
\includegraphics[width=0.9\columnwidth]{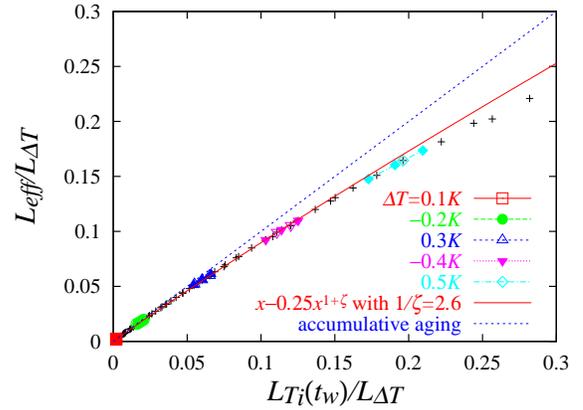}
\caption{Scaling plot of $\Leff$ with $\Lovlp= L_0 (a |\Delta T|/J)^{-1/\zeta}$ 
with $a=5$ and $1/\zeta=2.6$. The fit is due to the scaling ansatz
for the weakly perturbed regime explained in the text.
The open/filled symbols are for positive/negative T-shifts.
\label{plot}}
\end{figure}

Supposing that $\Lovlp > L_{T_{\rm i}}(\tw)$, all length scales smaller than the domain size $L_{T_{\rm i}}(\tw)$ belong to the weakly perturbed regime. 
A finite region of size $L$ within a domain can then become
out-of-equilibrium with the probability $p(L/\Lovlp)$. 
This reflects the breaking up of domains into smaller ones, yielding
a broad distribution of
domain sizes up to the maximum  $\Lmax=L_{T_{\rm i}}(\tw)$.
Then a simple scaling argument
gives that the effective domain size surrounding an arbitrary point of
the system on average reaches,
$L_{\rm eff}=\Lovlp F(\Lmax)/\Lovlp)$ with
$F(x)= x- c x^{1+\zeta}$.
Here, the 2nd term in $F(x)$, with $c$ being a positive constant
yields a correction term, which is the first order of $\tilde{\rho}(0)\Delta
U/\Upsilon$ to the no-chaos limit of $\Leff$.
Higher order terms can be neglected 
for small enough  $|\Delta T|$.
It can be seen in Fig.~\ref{plot} that the ansatz fits our data for
$|\Delta T| \lesssim 0.4$~K.

In the simulations of Berthier-Bouchaud \cite{bb-comment},
length scales $L \lesssim 5$ are probed  while $\Lovlp \gtrsim 20$.
In \cite{BB02} the spatial correlation function 
is found to decay to $0$ only after $L_{T_{i}}(\tw)$
being consistent with $\Lmax=L_{T_{i}}(\tw)$.
Then the rejuvenation found in \cite{bb-comment}
can be understood as partial rejuvenation due 
to the temperature chaos effect in a weakly perturbed 
regime. 
We note that 
emergence of the temperature chaos effect has been confirmed 
in the same model as in \cite{bb-comment} with modest system sizes \cite{HI}.
Indeed we have found that the ansatz proposed above also fits the 
result of Berthier-Bouchaud \cite{bb-comment} very well with 
$c=0.15$ and  $\zeta=1$. However, the largest $T$-shifts
used in \cite{bb-comment} involve temperatures in the range $0.8-0.9T_{\rm g}$,
within which  numerical (but not experimental) 
length/time scales are  dominated
by critical fluctuations as shown by Fig 1. of \cite{BB02}.

We conclude that the results reported in \cite{letter} 
can be consistently understood in terms of the temperature chaos effect in a weakly perturbed regime. 
We are not aware of 
any other concrete scenarios which predict the symmetrical effect 
of positive and negative T-shifts reported in \cite{letter}.

\acknowledgments
This work was financially supported by the Swedish Research Council
(VR).

P. E. J{\"o}nsson$^1$, H. Yoshino$^2$ and P. Nordblad$^1$

$^1${Department of Materials Science, Uppsala University, 
Box 534, SE-751 21 Uppsala, Sweden }

$^2${Department of Earth and Space Science, 
Osaka University, Toyonaka, 560-0043 Osaka, Japan}

\end{document}